\documentclass[11pt,twoside]{article}
\usepackage{FUSE2004}
\usepackage{natbib}

\usepackage{epsf}
\usepackage{psfig}
\usepackage{lscape}

\begin{document}

\title{FUSE's Five Years of Progress on the Interstellar Medium}
\author{JASON TUMLINSON} \affil{Department of
Astronomy and Astrophysics, University of Chicago, 5640 S. Ellis
Ave., Chicago, IL 60637}

\begin{abstract}
I review the five years of progress by {\it FUSE} on current
topics in the interstellar medium. {\it FUSE}'s sensitivity and
unique access to the far ultraviolet allow investigators to solve
problems in all phases of the interstellar medium. I describe {\it
FUSE}'s contributions in four major areas: 1) the Local
Interstellar Medium (LISM), 2) the hot phase (\ion{O}{6}), 3) the
cold phase (H$_2$), and 4) interstellar gas abundances. I devote
particular attention to the common themes of ISM phase
interactions and changes with metallicity. As a whole, these
results show that {\it FUSE} is the most powerful machine ever to
address problems of the ISM, and that {\it FUSE} points vividly to
the future of ISM studies in the Galaxy, Local Group, and beyond.
\end{abstract}

\section{Introduction}

This contribution reviews the fundamental advances {\it FUSE} has
made in our understanding of the interstellar medium (ISM). To get
a broad picture of recent progress, I surveyed all publications
based on {\it FUSE} data appearing in the refereed literature
between 1999 and 2004. I do not distinguish results from {\it
FUSE} PI Team observations and GO investigations, but the
PI-dominated schedule of the first three observing cycles is
clearly apparent. A review of the more than one hundred
ISM-related GO programs proves the great potential that awaits us
in the future. The present volume also contains many excellent
contributions too new to make my July 2004 cutoff. Related ISM
topics are addressed in other contributions on deuterium (Pettini,
H\'{e}brard, Linsky, Draine), high-velocity clouds (HVCs),
including O~VI (Sembach, Collins, Fox), O~VI in the Galactic disk
(Bowen) and the Magellanic Clouds (Howk), dust (Clayton), and
supernova remnants (SNRs; Sankrit).

As a whole, these studies show that in terms of {\em topical
scope} (the number of diverse topics addressed) and {\em
astrophysical range} (6 decades of distance, 8 decades of column
density, and 3 decades of metallicity), {\it FUSE} is the most
powerful instrument that has ever addressed problems of the ISM.
The first point follows from {\it FUSE}'s wavelength coverage and
efficient multiplexing, and the second from its unprecedented
sensitivity. These themes recur throughout our review of specific
results, and will be revisited in \S~8 to motivate future work.

In the following sections I discuss {\it FUSE}'s specific
contributions to four areas of ISM research. First, in \S~2 I
sketch out our basic understanding of interstellar processes to
help assess the basic character of {\it FUSE}'s contributions. In
\S~3 I review {\it FUSE} results on the structure and ionization
of the Local Interstellar Medium (LISM). Section 4 discusses {\it
FUSE} work on the hot phase of the ISM, as traced by O~VI.
Section~5 turns to the studies of H$_2$ in the Galaxy and
Magellanic Clouds. Section~6 briefly summarizes other results on
the Magellanic Clouds. In \S~7 I discuss the wide range of {\it
FUSE} interstellar abundance studies and their astrophysical
implications. Finally, the concluding section (\S~8) revisits the
themes introduced in \S~2 and considers how {\it FUSE}'s legacy
will inspire the conception and design of future FUV instruments.

\section{{\it FUSE} and the Cartoon ISM}

Figure 1 shows a simplified cartoon of interstellar ``ecology''
that will help us understand the unique contributions of {\it
FUSE}. Starting from the cold, dense molecular clouds where stars
are born, we proceed to the ``blister'' regions where young hot
stars emit the winds and strong UV radiation that heat and disrupt
their immediate environments. On Myr timescales, hot stars end
their lives as supernovae and inject energy and mass into the
ambient medium. Where hot-star formation occurs in coeval
clusters, multiple supernovae can evacuate large superbubbles
filled with hot ionized medium (HIM), some of which may escape
from the Galactic disk into the halo in a ``Galactic fountain''.
This hot gas eventually cools, recombines, returns to the disk,
and mixes with other phases to form the cold and warm neutral
media (CNM and WNM). These classical diffuse clouds, traced by H~I
and H$_2$, eventually cool below 100 K and coalesce to form new
molecular clouds. The entire process is thereafter repeated.

\begin{figure*}[!t]
\plotone{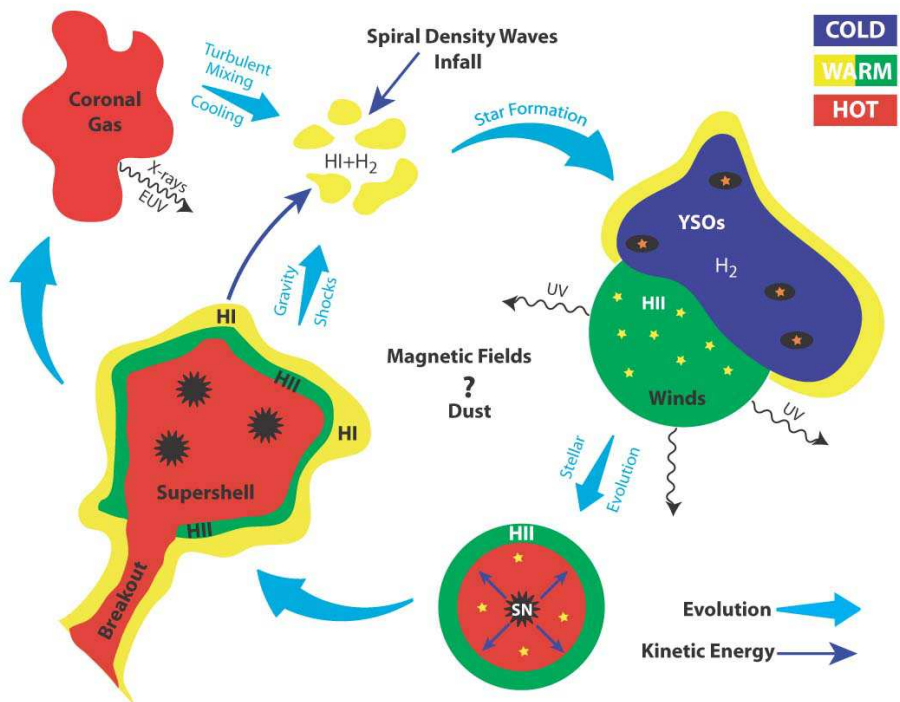} \caption{A cartoon of ISM ecology. {\it
FUSE} has added details to the ``evolution'' arrows and added a new
dimension -- metallicity -- to this figure.}
\end{figure*}

The relative levels of detail in our understanding of the stages
track closely our capability to detect and study their emission
and/or absorption. Because nearby examples are spatially resolved
and emit at optical wavelengths, young star-forming regions are
probably the best-understood objects in the diagram. By contrast,
the hotter (HIM) gas, which emits in X-rays and absorbs in the UV,
and the diffuse WNM and CNM, which emit in H~I and absorb in H~I,
H$_2$, and the low ions, are not as well understood.

{\it FUSE}'s fundamental contributions to ISM studies add both
sophistication and dimension to this simple picture. {\it FUSE}
provides details where we currently have only the speculative
``evolution'' arrows on the diagram. In particular, {\it FUSE}'s
unique access to the O~VI $\lambda\lambda$1032,1038 doublet has
enabled study of the highly-ionized boundary regions where the hot
and cold phases interact. {\it FUSE}'s access to the hundreds of
Lyman and Werner ro-vibrational lines of H$_2$ and extensive
abundance studies have added a metallicity dimension to this
diagram. Using {\it FUSE} we have shown that H$_2$ responds to the
metallicity and ambient radiation field in its environment, so we
are learning how interstellar processes change their character in
chemically primitive galaxies. FUSE's sensitivity and efficient
multiplexing have placed most of its substantive conclusions on
statistically sound footing.

\section{The Nature of the Local Interstellar Medium}

\begin{figure*}[!ht]
\plotfiddle{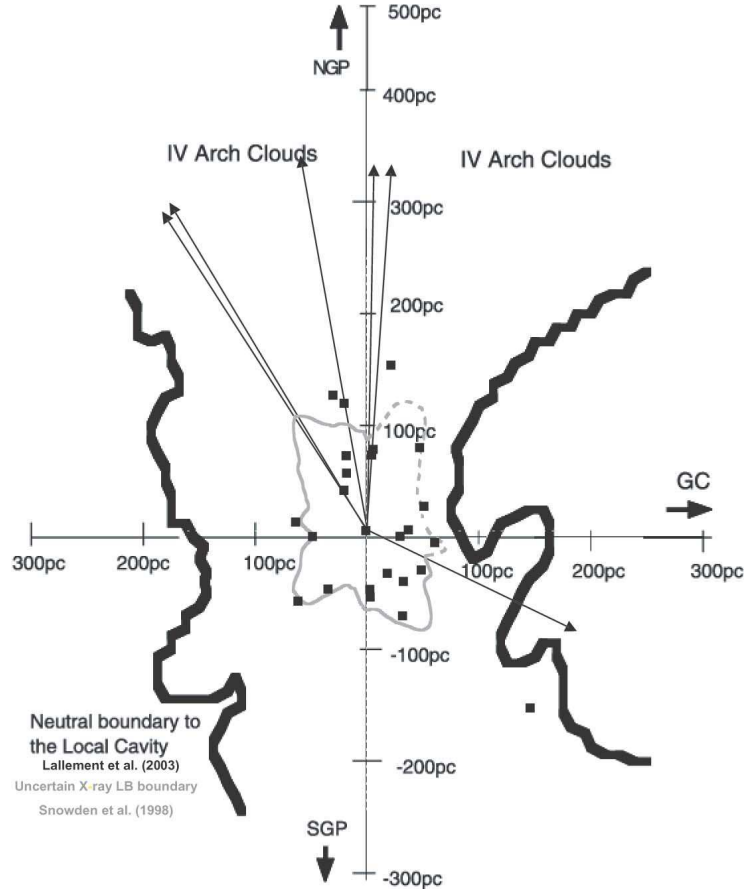}{4.0in}{0}{50}{50}{-150}{0}
\caption{The Local Bubble contains a hot, X-ray emitting plasma of
uncertain extent (X-ray contour in grey;
\citeauthor{1998ApJ...493..715S}~\citeyear{1998ApJ...493..715S}).
The neutral boundary to the Local Cavity is marked with the heavy
black contour
(\citeauthor{2003A&A...411..447L}~\citeyear{2003A&A...411..447L},
figure from
\citeauthor{2004A&A...414..261W}~\citeyear{2004A&A...414..261W}).
The grey squares mark the WD sample observed by FUSE (Oegerle et al.
2004, in prep). The grey arrows trace the complete sightlines
observed in O VI emission. \label{lism-figure}}
\end{figure*}

The Solar System resides in the Local Interstellar Cloud (LIC),
one of the Local Interstellar Medium Clouds (LISM), all embedded
within the Local Bubble (LB). The LB is filled with hot diffuse
gas and which apparently extends into the low Galactic halo
(Figure~\ref{lism-figure}). In a twist to the truism that higher
sensitivity pushes outward, {\it FUSE} has opened new windows into
the LISM by observing faint white dwarfs (WDs) at $< 100$ pc from
the Sun, generally closer than the bright O-stars accessible to
{\it Copernicus}. Using measurements of Ar I, N I/II/III, and O I
toward these new targets, \citet{2000ApJ...538L..81J} inferred
that the LISM clouds are steadily photoionized by EUV from OB
stars and hot gas rather than ``overionized'' from an earlier
period of strong heating by SNe. The \citet{2002ApJS..140...81L}
survey of 31 WDs (squares in Figure~\ref{lism-figure}) confirmed
this result, found little H$_2$ in the LISM, and measured the
local C~II cooling rate.

In studies of the hot phase of the local ISM,
\citet{2001ApJ...560..730S}, \citet{2001ApJ...552L..69D}, and
\citet{2002A&A...394..691W} detected O~VI emission along 6
high-latitude sightlines (marked by arrows in
Figure~\ref{lism-figure}). Using a ``shadowing'' technique that
observed an optically thick cloud just beyond the LB boundary,
\cite{2003ApJ...589..261S} placed a strict upper limit on O~VI
emission from the LB itself. This limit implies that LB and LISM
interfaces are rare, small, and/or produce little O~VI. The inferred
cloud sizes of $\sim$ 10 pc at $T =$ 300,000 K suggest conductive
interfaces or mixing layers between hot and cold gas.

These results have been confirmed by the larger O~VI absorption
survey by Oegerle et al. (2004, in preparation), who found weak or
absent O~VI absorption toward the local WD sample. This surprising
result suggests that patchy O~VI occurs only where conductive
interfaces between the hot LB gas and cool LISM clouds are not
quenched by magnetic fields.

{\it FUSE} has resolved the debate over the source of ionization
in favor of photoionization models rather than incomplete recovery
from a past highly ionized condition, as had been proposed. {\it
FUSE} has also thoroughly characterized the distribution of O
VI-bearing hot gas in the LISM, implying that the interfaces
between the hot LB gas and the nearby neutral ISM are more
complicated than previously believed. The theme of ISM phase
interaction continues into the next section, where I review O VI
results from the more distant ISM.

\section{O VI and Phase Interactions}

{\it FUSE}'s access to the O VI $\lambda\lambda$1032,1038 doublet
provides us with our clearest window into the hot phase of the
interstellar medium. In an attempt to explain how neutral ISM
clouds could remain stable at high Galactic latitudes,
\citet{1956ApJ...124...20S} first predicted the existence of hot
interstellar gas extending to several kpc above the Galactic
plane. {\it Copernicus} first confirmed this prediction with short
Galactic disk sightlines, but with its high sensitivity and
efficient multiplexing, {\it FUSE} has vindicated Spitzer's idea
in spectacular fashion. However, by studying more than 10 times as
many complete sightlines through the disk and halo, {\it FUSE} has
revealed that the distribution and character of the O~VI is
somewhat more complicated than Spitzer foresaw.

Surveys by \citet{2003ApJS..146....1W},
\citet{2003ApJS..146..125S}, \citet{2003ApJS..146..165S}, and
\citet{2003ApJ...586.1019Z} found four key features of the O~VI,
listed here with their physical implications.\footnote{Disk and
LMC/SMC O~VI results are discussed in the papers by Bowen and
Howk, respectively.} The O~VI is:

{\it Ubiquitous:} O VI is detected in 100 of 102 complete
sightlines through the Galactic thick disk and halo, so it traces
a common phenomenon in the ISM.

{\it Short-lived:} Because O VI reaches its peak ionization
fraction at 300,000~K, where solar-metallicity gas cools in $\leq
10^7$ yr, the O VI must trace short-lived, non-equilibrium
processes.

{\it Variable:} The strong 2--4 $\times$ variations in $N$(O~VI)
of over 10 pc –-- 1 kpc \citep{2002ApJ...572..264H} imply that the
O VI arises in small structures rather than a stably stratified,
stable hot layer.

{\it Poorly correlated:} The poor correlation of $N$(O~VI) with H~I
(CNM), H$\alpha$ (WIM), and soft X-rays (HIM) suggests that the
transition temperature gas traced by O~VI lies outside these phases.

Conductive interfaces, turbulent mixing layers, radiative cooling
zones, supernova remnants, and a Galactic fountain flow all meet
the basic criteria. Eight sightlines with {\it HST} data on C~IV,
N~V, and O~VI ($T =$ 1, 2, 3 $\times$ 10$^5$ K) favor a cooling
Galactic fountain model with $\dot{M} \sim 1.4$ M$_{\odot}$
yr$^{-1}$ from either side of the disk, but all these
non-equilibrium transition zones probably contribute at some level
\citep{2004ApJ...605..205I,2004ApJ...607..309I}. In the future we
need spectral resolution $R \sim$ 100,000 to "count" individual
interfaces, more sightlines with supporting data on C~IV and N~V
(a perfect project for the {\it Cosmic Origins Spectrograph}), and
new theory to help isolate the different ionization mechanisms.
Thanks to its unique waveband and efficient multiplexing, {\it
FUSE} has made a major advance in the understanding of how the hot
and cool phases of the ISM interact.

\section{H$_2$ and the Low-Metallicity ISM}

Although H$_2$ was shown by {\em Copernicus} to be widespread in
the Galactic disk, {\it FUSE} has found diffuse H$_2$ virtually
everywhere it has looked, including:

{\it The Galactic disk and halo}, where it forms a baseline for
comparison to other environments (Shull et al. 2004, in
preparation; Gillmon contribution);

{\it Galactic intermediate velocity clouds}, where it links them
closely to the Galactic disk \citep{2003ApJ...586..230R};

{\it The Monoceros Loop SNR}, where it may have reformed behind
the remnant's shock front \citep{2002A&A...381..566W};

{\it High-velocity clouds}, where H$_2$ formed {\it in situ} shows
that HVCs can have significant dust \citep{2001ApJ...562L.181R};

{\it The Small and Large Magellanic Clouds}, where it reflects low
metallicity and robust star formation
\citep{2001A&A...379...82B,2002ApJ...566..857T};

{\it The Magellanic Stream and Bridge}, showing that it can survive
tidal stripping from dwarf galaxies \citep{2001AJ....121..992S,
2002ApJ...578..126L}; and

{\it The Local Group spiral galaxy M33}, where it indicates how to
analyze composite spectra of many sources
\citep{2003A&A...398..983B}.

Thanks to these surprising results we are now learning how to use
H$_2$ as a sensitive indicator of local physical conditions, such
as temperature, density, metallicity, and radiation field. Two
such studies are reviewed here, while a large number of
contributed talks and posters in this volume present additional
H$_2$ and CO results.

Richter et al. (2003) detected H$_2$ in 14 of 61 H~I IVCs
($|v_{LSR}| = 30 - 90$ km s$^{-1}$). Because H$_2$ dissociation by
FUV radiation occurs in less than 10$^3$ yr where there is no
competing molecule formation on the surfaces of dust grains, the
widespread H$_2$ in the IVCs must have formed {\it in situ}.
Kinematically the IVCs appear to be Galactic fountain gas
returning to the disk in the form of small ($\sim$0.1 pc), dense
($n_H \sim 30$ cm$^{-3}$) clouds. That {\it FUSE} has identified
both the beginning and end of the Galactic fountain using
essentially the same set of sightlines illustrates {\it FUSE}'s
multiplexing power.

\begin{figure*}[!ht]
\plotone{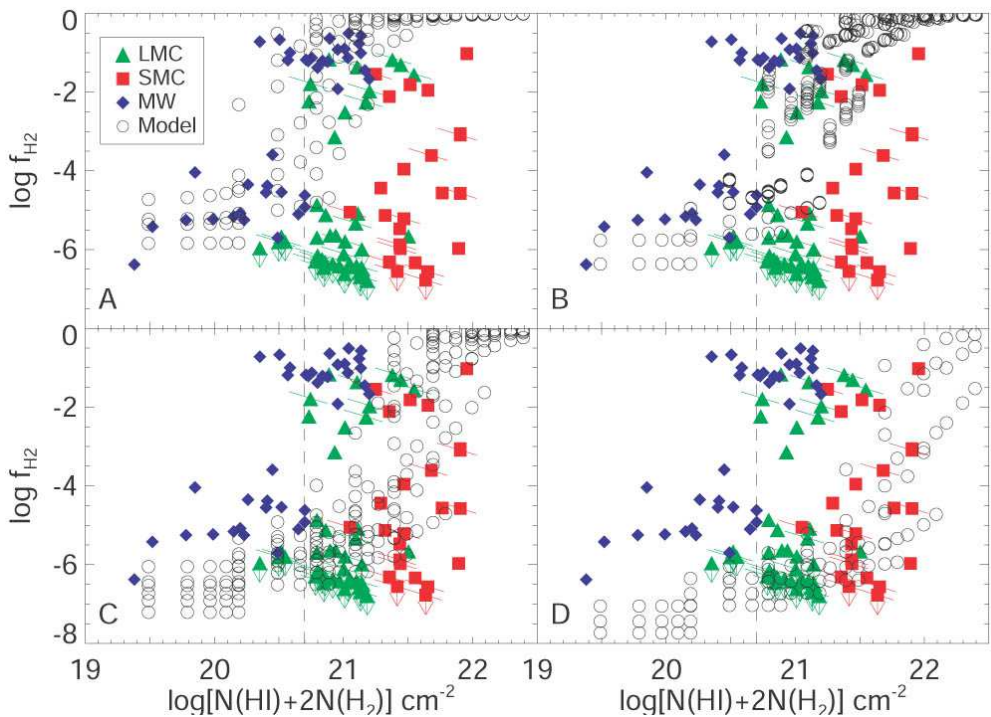} \caption{$H_2$ abundances relative to
total H in the Galactic disk (diamonds), LMC (triangles), and SMC
(squares) from \citet{2002ApJ...566..857T}. Models with Galactic
radiation field and grain formation rate match the Galactic points
in panel A. Both elevated radiation fields and low grain formation
rate per H atom are required to explain the low abundances seen in
the SMC and LMC (panel D).\label{tumlinson-lmc-fig}}
\end{figure*}

In a survey of 70 LMC/SMC sightlines, \citet{2002ApJ...566..857T}
found evidence of elevated FUV radiation fields and reduced H$_2$
grain formation rates in the LMC and SMC.
Figure~\ref{tumlinson-lmc-fig} shows the Magellanic Cloud data and
models for Galactic conditions in panel A and 10 -- 100$\times$
Galactic radiation field and 1/3 -- 1/10 Galactic grain formation
rate in panel D. These data confirm that H$_2$ formation-destruction
balance shifts with metallicity and starburst activity, as predicted
by theoretical models~\citep*{2003ApJ...582..810B}. Molecular
hydrogen is therefore providing critical insights into how
interstellar processes change with metallicity, and may also allow
us to diagnose physical conditions and stellar populations where
they are not resolved, or even detected (such as in damped
Ly$\alpha$ systems).

\section{Magellanic ISM}

{\it FUSE} investigators have done fundamental work on the LMC and
SMC ISM, providing a critical link between the Milky Way and
high-redshift galaxies on the cosmic "metallicity ladder". Some
examples of {\it FUSE} LMC/SMC work are the dust extinction curve
study by \citet{2001PASP..113.1205H}, the
\citet{2002ApJ...578..126L} study of abundances, ionization, and
molecules in the Magellanic Bridge, and the extremely thorough
sightline analysis of the SMC star Sk 108 by
\citet{2003ApJS..147..265M}. In studies of Magellanic CO,
\citet{2001A&A...379...82B} and \citet{2004A&A...422..483A} find
CO/H$_2$ ratios that match the Galaxy, despite low $Z$. A broad
view of the LMC and SMC interstellar medium is provided by the
\citet{2002ApJS..139...81D} atlas, which is a vital starting point
for the analysis of LMC/SMC data or for the planning of new {\it
FUSE} observations. Thus {\it FUSE} is beginning to make good on
the long-standing promise of using the Magellanic Clouds as a
template for understanding damped Ly$\alpha$ systems and
primordial galaxies.

\section{{\it FUSE} Abundances Unlock Interstellar Secrets}

In accord with our theme of adding a metallicity dimension to the
interstellar ecology diagram, {\it FUSE} has obtained interstellar
abundances over 6 decades of distance and 3 decades of metallicity
(Figure~\ref{abundance-figure}). {\it FUSE} can measure abundances
directly for elements with dominant ions in the FUV and assist
studies of many other elements by accurately determining $N$(H~I)
and $N$(H$_2$). As seen below, in almost every case abundances
measured or assisted by {\it FUSE} reveal key insights into the
operation of physical processes at low metallicity or high
density. Some examples are:

\begin{figure*}[!t]
\plotone{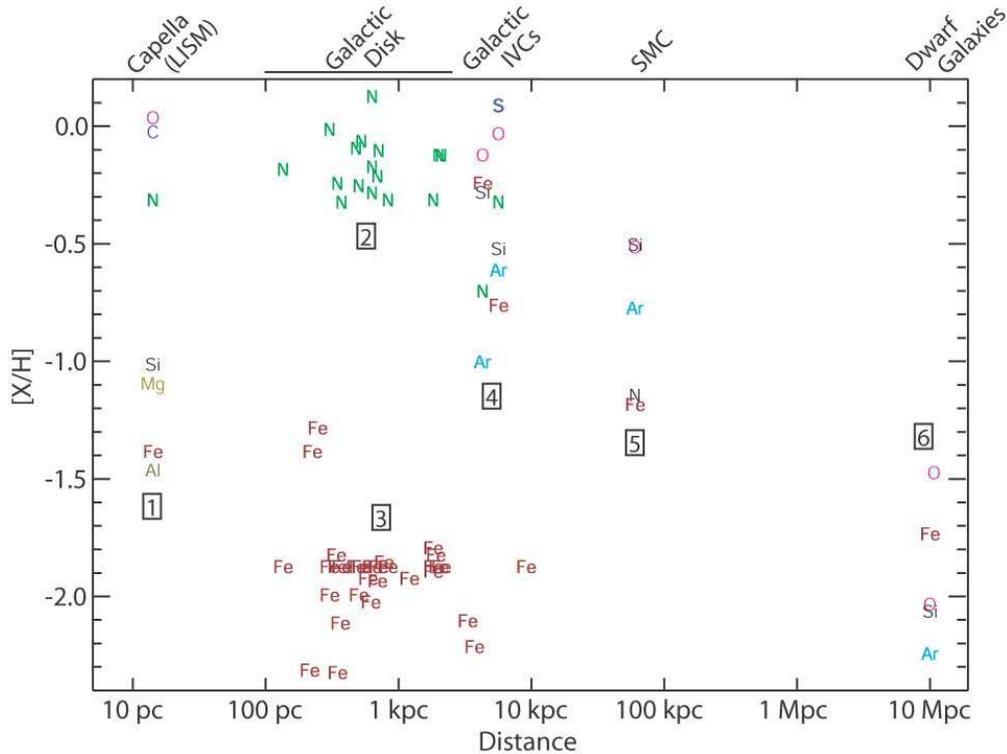} \caption{Interstellar abundances in
diverse environments. {\it FUSE} has measured details abundances
from the 14 pc Capella sightline to the metal-poor ISM of I Zw 18 at
10 Mpc. The boxed numerals correspond to the summary points in
\S~7.\label{abundance-figure}}
\end{figure*}

\begin{itemize}
\item[1.] Wood et al. (2002) measured roughly solar CNO but 10$\times$
depleted Si, Mg, Fe, Al, which indicates significant dust on {\it
FUSE}'s shortest sightline.

\item[2.] \citet{2003ApJ...596L..51K} found strong
variations in the interstellar N~I abundance in the Galactic disk.
A deficit of N~I at high $N$(H) suggests a ``missing N problem''
or poor understanding of N chemistry in dense clouds.

\item[3.] \citet{2002ApJ...573..662S} found that Fe depletions in
``translucent cloud" sightlines do not increase beyond $A_V \sim
1$. This finding supports the conclusion from their
high-extinction H$_2$ survey \citep{2002ApJ...577..221R} that true
translucent clouds, if they exist, have not yet been found.

\item[4.] \citet{2001ApJ...549..281R,2001ApJ...559..318R} measured
near-solar abundances for non-refractory elements (S, O) in IVCs,
suggesting that this extra-planar gas originates in the Galactic
disk.

\item[5.] \citet{2003ApJS..147..265M} found undepleted Ar, O, S, and
Si, but depleted Fe toward Sk 108 in the SMC, which suggests unusual
grain composition or lower dust-to-metal ratios at 40\% solar
metallicity.

\item[6.] \citet{2003ApJ...595..760A} used abundances in the blue
compact dwarf galaxy I~Zw~18 to diagnose its star formation
history. They found abundances consistent with ancient star
formation $> 1$ Gyr in the past (see her contribution).

\end{itemize}

In addition to abundances obtained directly from {\it FUSE} data,
{\it FUSE} contributes $N$(H~I) and/or $N$(H$_2$) to measurements
of $N$(H). This subtle capability of {\it FUSE} is especially
important for dense interstellar environments where the H$_2$
fraction is large. These studies have measured abundances for Kr
in the Galactic disk \citep{2003ApJ...597..408C}, refined O/H in
the Galactic \citep{2003ApJ...591.1000A}, and extensively probed
elements from C to Ge toward reddened disk stars
\citep{2002ApJ...576..241S,2003ApJ...596..350S}.

The abundance studies of distant objects (LMC, SMC, I~Zw~18)
illustrate all our major themes - the changes of interstellar
processes with metallicity, the push to new environments, and the
future potential of new instruments.

\section{Discussion, Conclusions, and the Future}

My literature review of {\it FUSE} results led me to four
conclusions about the present state and future prospects of
interstellar medium studies:
\begin{itemize}

\item[1.] {\it FUSE} has shifted our focus from characterizing
the basic properties of the ISM phases to detailed studies of
their interactions in transition zones. I hope that our poor
understanding of these regions and the high-quality database from
FUSE will motivate theorists to address this problem.

\item[2.] Thanks to {\it FUSE}, we are moving beyond simple
characterizations of H$_2$ in the Galactic disk and learning how
to use H$_2$ as a sensitive indicator of local physical conditions
in low-metallicity ISM.

\item[3.] To interstellar astrophysics, {\it FUSE} will be
remembered as the mission that extended ISM studies to external
galaxies, showing us a tantalizing glimpse of what awaits {\it
FUSE} and its successors.

\item[4.] Extending the detailed LMC/SMC studies by {\it FUSE}
throughout the Local Group and beyond should be the primary goal
of an FUV successor mission. The desired sensitivity will also
open new Galactic windows.

\end{itemize}

In the final analysis, we see that {\it FUSE} has made fundamental
contributions to our understanding of the ISM, in ways that no other
instrument can claim. These achievements derive directly from {\it
FUSE}'s high sensitivity and unique waveband. Thus {\it FUSE}
demonstrates the extraordinary potential of a successor with higher
sensitivity and resolution to make another qualitative leap in our
understanding. Indeed, {\it FUSE} shows that any future mission
focussed on ISM and star formation studies should cover the FUV to
get access to the important tracers discussed here. {\it FUSE} has
also provided an excellent science case by leading us out of the
narrow confines of the Galactic disk into diverse interstellar
environments. This is how {\it FUSE} will be remembered to
interstellar astronomers of the future.


\begin{thebibliography}{}

\bibitem[Aloisi et al.(2003)]{2003ApJ...595..760A} Aloisi, A., Savaglio,
S., Heckman, T.~M., Hoopes, C.~G., Leitherer, C., \& Sembach, K.~R.\
2003, \apj, 595, 760

\bibitem[Andr{\' e} et al.(2003)]{2003ApJ...591.1000A} Andr{\' e}, M.~K.,
et al.\ 2003, \apj, 591, 1000

\bibitem[Andr{\' e} et al.(2004)]{2004A&A...422..483A} Andr{\' e}, M.~K.,
et al.\ 2004, \aap, 422, 483

\bibitem[Bluhm \& de Boer(2001)]{2001A&A...379...82B} Bluhm, H.~\& de Boer,
K.~S.\ 2001, \aap, 379, 82

\bibitem[Bluhm et al.(2003)]{2003A&A...398..983B} Bluhm, H., de Boer,
K.~S., Marggraf, O., Richter, P., \& Wakker, B.~P.\ 2003, \aap, 398,
983

\bibitem[Browning, Tumlinson, \& Shull(2003)]{2003ApJ...582..810B}
Browning, M.~K., Tumlinson, J., \& Shull, J.~M.\ 2003, \apj, 582,
810

\bibitem[Cartledge, Meyer, \& Lauroesch(2003)]{2003ApJ...597..408C}
Cartledge, S.~I.~B., Meyer, D.~M., \& Lauroesch, J.~T.\ 2003, \apj,
597, 408

\bibitem[Danforth et al.(2002)]{2002ApJS..139...81D} Danforth, C.~W., Howk,
J.~C., Fullerton, A.~W., Blair, W.~P., \& Sembach, K.~R.\ 2002,
\apjs, 139, 81

\bibitem[Dixon et al.(2001)]{2001ApJ...552L..69D}
Dixon, W.~V.~D., Sallmen, S., Hurwitz, M., \& Lieu, R.\ 2001, \apjl,
552, L69

\bibitem[Ferlet et al.(2000)]{2000ApJ...538L..69F} Ferlet, R., et al.\
2000, \apjl, 538, L69

\bibitem[Fox et al.(2003)]{2003ApJ...582..793F} Fox, A.~J., Savage, B.~D.,
Sembach, K.~R., Fabian, D., Richter, P., Meyer, D.~M., Lauroesch,
J., \& Howk, J.~C.\ 2003, \apj, 582, 793


\bibitem[Hoopes et al.(2002)]{2002ApJ...569..233H} Hoopes, C.~G., Sembach,
K.~R., Howk, J.~C., Savage, B.~D., \& Fullerton, A.~W.\ 2002, \apj,
569, 233

\bibitem[Howk et al.(2002)]{2002ApJ...572..264H} Howk,
J.~C., Savage, B.~D., Sembach, K.~R., \& Hoopes, C.~G.\ 2002, \apj,
572, 264

\bibitem[Howk et al.(2002)]{2002ApJ...569..214H} Howk, J.~C., Sembach,
K.~R., Savage, B.~D., Massa, D., Friedman, S.~D., \& Fullerton,
A.~W.\ 2002, \apj, 569, 214

\bibitem[Hutchings \& Giasson(2001)]{2001PASP..113.1205H} Hutchings,
J.~B.~\& Giasson, J.\ 2001, \pasp, 113, 1205

\bibitem[Indebetouw \& Shull(2004a)]{2004ApJ...605..205I} Indebetouw, R.~\&
Shull, J.~M.\ 2004, \apj, 605, 205

\bibitem[Indebetouw \& Shull(2004b)]{2004ApJ...607..309I} Indebetouw, R.~\&
Shull, J.~M.\ 2004, \apj, 607, 309

\bibitem[Jenkins et al.(2000)]{2000ApJ...538L..81J} Jenkins, E.~B., et al.\
2000, \apjl, 538, L81

\bibitem[Knauth et al.(2003)]{2003ApJ...596L..51K}
Knauth, D.~C., Andersson, B.-G., McCandliss, S.~R., \& Moos, H.~W.\
2003, \apjl, 596, L51


\bibitem[Lallement et al.(2003)]{2003A&A...411..447L} Lallement, R., Welsh,
B.~Y., Vergely, J.~L., Crifo, F., \& Sfeir, D.\ 2003, \aap, 411, 447

\bibitem[Lehner(2002)]{2002ApJ...578..126L} Lehner, N.\ 2002, \apj, 578,
126

\bibitem[Lehner et al.(2002)]{2002ApJS..140...81L} Lehner, N., Gry, C.,
Sembach, K.~R., H{\' e}brard, G., Chayer, P., Moos, H.~W., Howk,
J.~C., \& D{\' e}sert, J.-M.\ 2002, \apjs, 140, 81

\bibitem[Mallouris(2003)]{2003ApJS..147..265M} Mallouris, C.\ 2003, \apjs,
147, 265

\bibitem[Rachford et al.(2002)]{2002ApJ...577..221R} Rachford, B.~L., et
al.\ 2002, \apj, 577, 221

\bibitem[Richter et al.(2001a)]{2001ApJ...549..281R} Richter, P., Savage,
B.~D., Wakker, B.~P., Sembach, K.~R., \& Kalberla, P.~M.~W.\ 2001,
\apj, 549, 281

\bibitem[Richter et al.(2001b)]{2001ApJ...562L.181R}
Richter, P., Sembach, K.~R., Wakker, B.~P., \& Savage, B.~D.\ 2001,
\apjl, 562, L181

\bibitem[Richter et al.(2001c)]{2001ApJ...559..318R} Richter, P., Sembach,
K.~R., Wakker, B.~P., Savage, B.~D., Tripp, T.~M., Murphy, E.~M.,
Kalberla, P.~M.~W., \& Jenkins, E.~B.\ 2001, \apj, 559, 318

\bibitem[Richter et al.(2003)]{2003ApJ...586..230R}
Richter, P., Wakker, B.~P., Savage, B.~D., \& Sembach, K.~R.\ 2003,
\apj, 586, 230

\bibitem[Savage et al.(2003)]{2003ApJS..146..125S} Savage, B.~D., et al.\
2003, \apjs, 146, 125

\bibitem[Sembach et al.(2001)]{2001AJ....121..992S}
Sembach, K.~R., Howk, J.~C., Savage, B.~D., \& Shull, J.~M.\ 2001,
\aj, 121, 992

\bibitem[Sembach et al.(2003)]{2003ApJS..146..165S} Sembach, K.~R., et al.\
2003, \apjs, 146, 165


\bibitem[Shelton et al.(2001)]{2001ApJ...560..730S} Shelton, R.~L., et al.\
2001, \apj, 560, 730

\bibitem[Shelton(2003)]{2003ApJ...589..261S} Shelton, R.~L.\ 2003, \apj,
589, 261

\bibitem[Snow, Rachford, \& Figoski(2002)]{2002ApJ...573..662S} Snow,
T.~P., Rachford, B.~L., \& Figoski, L.\ 2002, \apj, 573, 662

\bibitem[Snowden et al.(1998)]{1998ApJ...493..715S} Snowden, S.~L., Egger,
R., Finkbeiner, D.~P., Freyberg, M.~J., \& Plucinsky, P.~P.\ 1998,
\apj, 493, 715

\bibitem[Sonnentrucker et al.(2002)]{2002ApJ...576..241S} Sonnentrucker,
P., Friedman, S.~D., Welty, D.~E., York, D.~G., \& Snow, T.~P.\
2002, \apj, 576, 241

\bibitem[Sonnentrucker et al.(2003)]{2003ApJ...596..350S} Sonnentrucker,
P., Friedman, S.~D., Welty, D.~E., York, D.~G., \& Snow, T.~P.\
2003, \apj, 596, 350

\bibitem[Spitzer(1956)]{1956ApJ...124...20S} Spitzer, L.~J.\ 1956, \apj,
124, 20

\bibitem[Tumlinson et al.(2002)]{2002ApJ...566..857T} Tumlinson, J., et
al.\ 2002, \apj, 566, 857

\bibitem[Wakker et al.(2003)]{2003ApJS..146....1W} Wakker, B.~P., et al.\
2003, \apjs, 146, 1

\bibitem[Welsh, Rachford, \& Tumlinson(2002)]{2002A&A...381..566W} Welsh,
B.~Y., Rachford, B.~L., \& Tumlinson, J.\ 2002, \aap, 381, 566

\bibitem[Welsh et al.(2002)]{2002A&A...394..691W} Welsh, B.~Y., Sallmen,
S., Sfeir, D., Shelton, R.~L., \& Lallement, R.\ 2002, \aap, 394,
691

\bibitem[Welsh, Sallmen, \& Lallement(2004)]{2004A&A...414..261W} Welsh,
B.~Y., Sallmen, S., \& Lallement, R.\ 2004, \aap, 414, 261

\bibitem[Wood et al.(2002)]{2002ApJ...581.1168W} Wood,
B.~E., Redfield, S., Linsky, J.~L., \& Sahu, M.~S.\ 2002, \apj, 581,
1168

\bibitem[Zsarg{\' o} et al.(2003)]{2003ApJ...586.1019Z}
Zsarg{\' o}, J., Sembach, K.~R., Howk, J.~C., \& Savage, B.~D.\
2003, \apj, 586, 1019

\end{thebibliography}
\end{document}